# An explanation of the distribution of inter-seizure intervals

M.V. Simkin and V.P. Roychowdhury
*Department of Electrical Engineering, University of California, Los Angeles, CA 90095-1594*

Recently Osorio *et al* [1] reported that probability distribution of intervals between successive epileptic seizures follows a power law with exponent 1.5. We theoretically explain this finding by modeling epileptic activity as a branching process, which we in turn approximate by a random walk. We confirm the theoretical conclusion by numerical simulation.

Recently Osorio *et al* [1] reported that probability distribution of epileptic seizure energies and inter-seizure intervals follow a power law with exponents 1.67 and 1.5 respectively. Earlier Beggs and Plenz [2] had observed spontaneous neuronal avalanches in neocortical tissues. The distribution of sizes of these avalanches, computed by summing local field potentials, followed a power law with exponent 1.5 as in critical branching process. The relevance of the branching process to explanation of seizure energies distribution is thus obvious. Here we show that in addition the distribution of inter-seizure intervals is also consistent with the branching process.

Epileptic seizures result from simultaneous firing of large number of neurons in the brain. Anninos and Cyrulnik [3] proposed a neural net model for epilepsy a simplified version of which we will use in this study. To start with, we introduce time discretization. After a neuron has fired, it cannot fire again for a time interval known as refractory period. Therefore, the minimum interval between the beginnings of two subsequent firings of a neuron is the sum of spike duration and refractory period. This interval is few milliseconds [4]. We will use this interval as our time unit. Suppose that at given time step *N* neurons are firing. How many neurons will be firing next time step? Consider one of these firing neurons. Its axon connects to synapses of thousands of neurons. Some of them are almost ready to fire: their membrane potential is close to the firing threshold and one impulse from our neuron will be sufficient to surpass this threshold. The aforementioned experiment of Beggs and Plenz [2] suggest that the average number $\lambda$ of ready to fire neurons among those to which our neuron is connected is very close to unity. Only in this case we get a critical branching process and a power law distribution of avalanches. We have many neurons that possibly can be induced to fire by firing of our neuron and probability for each of them to fire is very small. Thus, the number of induced firings is Poisson distributed with mean $\lambda$ (which is very close to unity). The variance of the Poisson distribution equals its mean and therefore it also equals $\lambda$. If at given time step a large number *N* of neurons are firing then the number of neurons firing the next time step will come from a normal distribution with mean $\lambda N$ and variance $\lambda N$. In addition to induced firings, some neurons will fire spontaneously. We assume that the number of spontaneously firing neurons at each time step comes from a Poisson distribution with mean *p*. The change in the number of firing neurons is

$$\Delta N = (\lambda - 1)N + p + \sqrt{N}z \qquad (1)$$

where *z* is a normally distributed random number with zero mean and unit variance. The number of firing neurons, *N*, performs a random walk, with the size of the step proportional to $\sqrt{N}$. Eq.(1) can be simplified by changing variable from *N* to $x = \sqrt{N}$. Using Ito's formula [5], we get

$$\Delta x = \frac{(\lambda-1)}{2}x - \left(\frac{1}{8} - \frac{p}{2}\right) \times \frac{1}{x} + \frac{1}{2}z \qquad (2)$$

In the limit of large *x*, the term, inversely proportional to *x*, can be neglected in the first approximation. When $\lambda$ is very close to unity the first term can also be neglected. Equation (2) reduces to $\Delta x = 1/2\, z$ which means that $\sqrt{N}$ performs a simple random walk. A well know result in random walk theory is that the distribution of first return times to zero (or to any other chosen point) follows a power law with exponent $3/2$[1]. In the experiment [1], seizures were counted when electric intensity of epileptic discharges reached certain threshold. This in our model is equivalent to $\sqrt{N}$ reaching certain threshold. Then the distribution of first returns into seizure (inter-seizure intervals) is the same as the distribution of random walk's return times.

Now let us study Equation (2) without neglecting any terms. In the particular case $p = 1/4$ the second term cancels out and in the case $\lambda < 1$ we get a well studied problem of Brownian motion in a harmonic potential [7]. In the general $p$ case we get Brownian motion in the potential

$$U(x) = -\frac{(\lambda-1)}{4}x^2 + \left(\frac{1}{8} - \frac{p}{2}\right) \times \ln(x). \quad (3)$$

One can find probability density of $x$, by solving the corresponding Fokker-Planck equation. Alternatively it can be found as a Boltzmann distribution in the potential given by Eq.(3) at an appropriate temperature. The result is:

$$P(x) \sim \exp(-8U(x)) = \exp(2(\lambda-1)x^2)\frac{1}{x^{1-4p}} \quad (4)$$

The probability distribution of $N = x^2$ can be immediately found using Eq.(4)

$$P(N) \sim \exp(2(\lambda-1)N)\frac{1}{N^{1-2p}} \quad (5)$$

In the case when $p = 0$ and $\lambda = 1$ we get $P(N) \sim 1/N$. This one expects from the theory of branching processes [8]. The survival probability after $N$ generations for a critical branching process is $1/N$, while the expectation number for the number of individuals is 1. This means that the average size of surviving family is $N$, while the probability is $1/N$. In the case $p > 0$, branching processes overlap and this leads to a modified power law exponent. When $p = 0.5$ the power law cancels out and when $p > 0.5$ the power law exponent becomes positive. Experimentalists did not report the data which can be compared to Eq.(5) but it is most likely contained in their data files and they will be able to compare it with our theory. An important implication of Eq.(5) is that the random walk spends less time at higher values of $N$. This means that seizures are shorter than intervals between seizures.

One way to get a critical (or more precisely slightly subcritical) branching process would be to use the Self-organized criticality model [9]. The Sandpile model can be easily recast in neural network terms: accumulation of send grains corresponds to integration, toppling to firing, and spontaneous firing to adding sand grains. In practice it is easier to simulate a branching process than the generating it SOC system. Figures 1-3 show results of such simulation. The parameters were $\lambda = 1 - 10^{-8}$ and $p = 0.5$. The seizure threshold was set at $4 \times 10^8$ firing neurons. The seizure intensities were defined as total number of neuron firings during seizure. The system was simulated for $10^{11}$ time steps. Remember that time step is the sum of firing duration and refractory period. A reasonable estimate for this is two milliseconds. Thus, our simulation run corresponds to over six years. The longest inter-seizure interval was $5 \times 10^{10}$ time steps (over three years). The longest seizure was $10^7$ time steps (about five hours). The distribution of inter-seizure intervals is well described by an inverse power law with exponent 1.47. And the distribution of seizure intensities - by an inverse power law with exponent 1.48.

This research gives an answer to the old question "Do seizures beget seizures?" [10]. After symptoms of seizure end, there still remains for some time epileptic activity in the brain. It is easy for this activity to surpass the threshold again soon. If there was no seizure recently this most likely means that the epileptic activity is minimal or absent and it will take more time to build up the activity to pass the threshold. This model also gives an alternative explanation to power law distribution of intervals in other than epileptic human activity [11].

---
[1] One can get this by applying Stirling's formula to the result in chapter III.4 of Ref. [6]

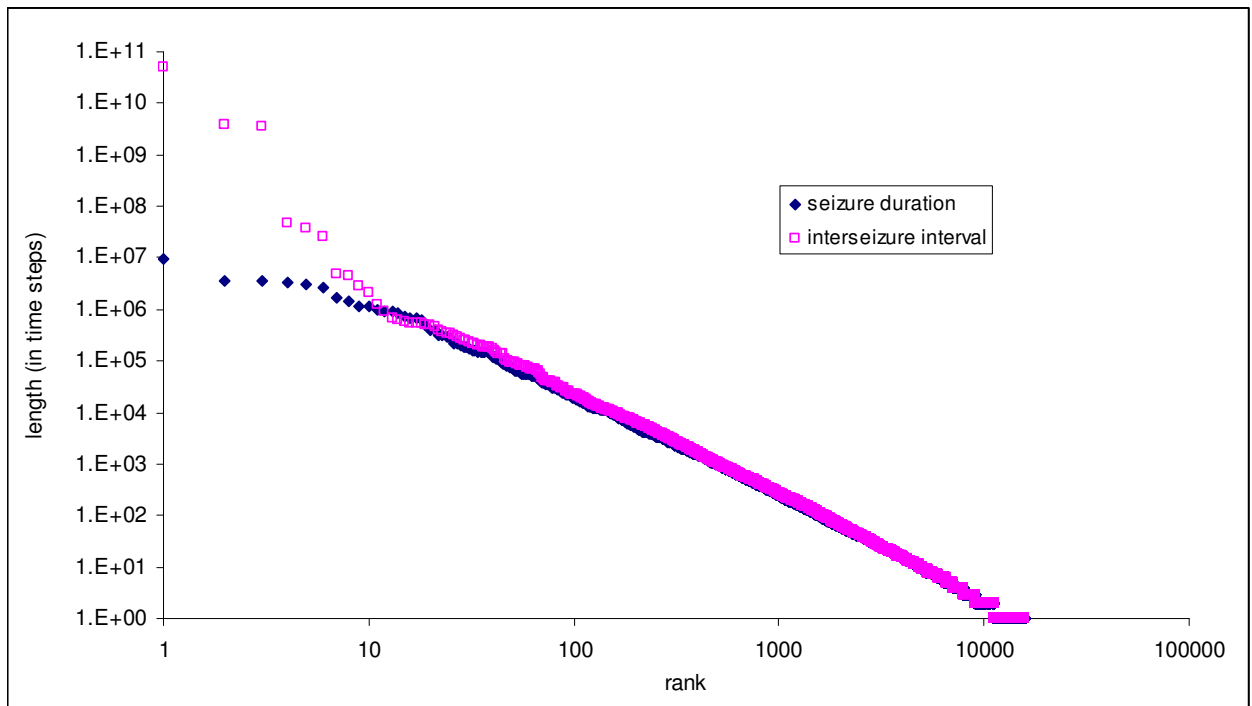

**Figure 1.** Results of numerical simulation of branching epileptic process model with $p = 0.5$ and $\lambda = 1 - 10^{-8}$. The simulation was run for $10^{11}$ time steps, which correspond to over six years. The longest inter-seizure interval is over three years. The longest seizure is about five hours.

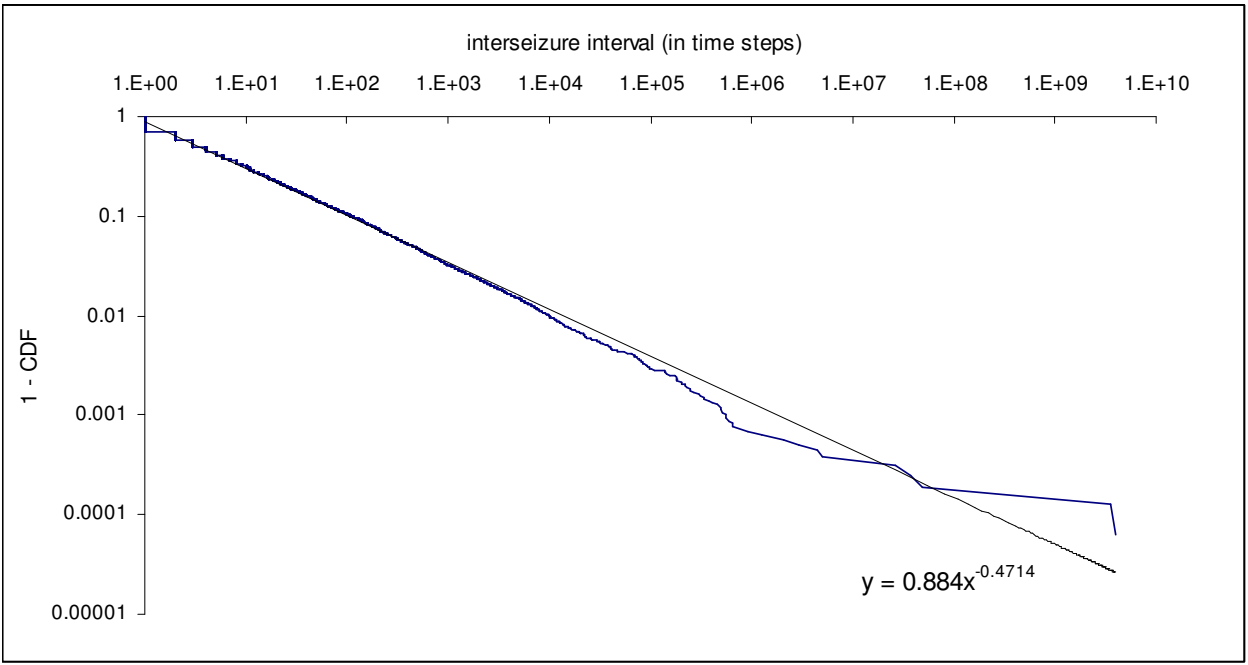

**Figure 2**. Cumulative distribution function (CDF) of inter-seizure intervals. 1 − CDF is fitted by a power law with exponent $-0.47$. This means that probability density function of inter-seizure intervals is a power law with exponent - $-1.47$.

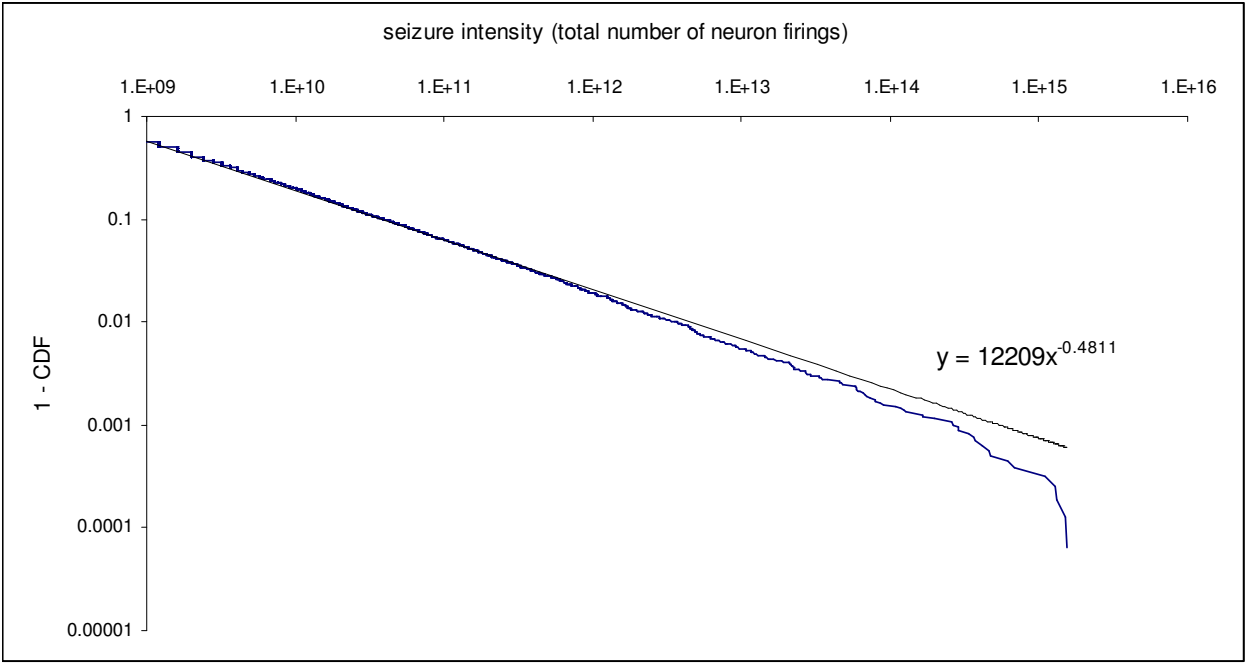

**Figure 3.** Cumulative distribution function (CDF) of seizure intensities. 1 − CDF is fitted by a power law with exponent $-0.48$. This means that probability density function of seizure intensities is a power law with exponent $-1.48$.